\documentclass[conference]{IEEEtran}
\IEEEoverridecommandlockouts

\usepackage{cite}
\usepackage[bookmarks=false,hidelinks]{hyperref}
\usepackage{amsmath,graphicx, cite, booktabs, subfig}
\usepackage{amsmath,amssymb,amsfonts}
\usepackage{algorithmic}
\usepackage{graphicx}
\usepackage{textcomp}
\usepackage{xcolor}
\def\BibTeX{{\rm B\kern-.05em{\sc i\kern-.025em b}\kern-.08em
    T\kern-.1667em\lower.7ex\hbox{E}\kern-.125emX}}
\newcommand{\modelnametitle}{MusicGen-Stem}
\newcommand{\modelname}{MusicGen-Stem }

\title{\modelnametitle: Multi-stem music generation and edition through autoregressive modeling}

\author{
\IEEEauthorblockN{Simon Rouard*}
\IEEEauthorblockA{\textit{Meta \& UMR STMS} \\ \textit{IRCAM-CNRS-Sorbonne Univ.}}
\and
\IEEEauthorblockN{Robin San Roman*}
\IEEEauthorblockA{\textit{Meta, FAIR, Univ. de} \\ \textit{Lorraine, CNRS, Inria, Loria}}
\and
\IEEEauthorblockN{Yossi Adi}
\IEEEauthorblockA{\textit{Meta \& Hebrew}  \\ \textit{Univ. of Jerusalem}}
\and
\IEEEauthorblockN{Axel Roebel}
\IEEEauthorblockA{\textit{UMR STMS, IRCAM-CNRS}\\ \textit{Sorbonne Univ.}}
}

\begin{document}
\maketitle
\def\thefootnote{*}\footnotetext{Equal contribution}

\begin{abstract}
While most music generation models generate a mixture of stems (in mono or stereo), we propose to train a multi-stem generative model with 3 stems (bass, drums and other) that learn the musical dependencies between them. To do so, we train one specialized compression algorithm per stem to tokenize the music into parallel streams of tokens. Then, we leverage recent improvements in the task of music source separation to train a multi-stream text-to-music language model on a large dataset. Finally, thanks to a particular conditioning method, our model is able to edit bass, drums or other stems on existing or generated songs as well as doing iterative composition (e.g. generating bass on top of existing drums). This gives more flexibility in music generation algorithms and it is to the best of our knowledge the first open-source multi-stem autoregressive music generation model that can perform good quality generation and coherent source editing. Code and model weights will be released and samples are available on \href{https://simonrouard.github.io/musicgenstem/}{\textcolor{blue}{simonrouard.github.io/musicgenstem}}.
\end{abstract}


\section{Introduction}
\label{sec:intro}
Recent models for music generation~\cite{musicgen, stableaudio, musiclm, tal2024joint} allow generating long and coherent audio sequences of up to several minutes with reasonable audio quality. Although recent studies provide rich conditions for controlling the generated music~\cite{tal2024joint, wu2024music}, the dominant approach is still text instructions. While these text prompts have the benefit of allowing extensive control over high-level musical characteristics of the generated music (like style, instrumentation, or mood), they remain limited with respect to precise control, specifically editing. For example, the production of a drum track for a given music piece can not be achieved using only a textual description of the result. 

These limitations have led to research activities that aim extending the use cases for music generation models. One line of research addresses the generation of musically coherent stems (or tracks) conditioned on audio input \cite{singsong, stemgen, jen1composer, multidiffusion_sep}. If successful, these approaches would provide innovative means for the generation of musical accompaniment, for the iterative, stem-wise creation of a musical piece, and in combination with source separation, may allow replacing a stem in a given musical piece.  

In line with these recent research activities the present work introduces MusicGen-Stem, an extension of \cite{musicgen} towards multi-stem music generation that is able to perform at the same time text-to-music generation, text and audio conditioned stem generation,  as well as iterative stem-by-stem generation. One of the main problems for training music generation for individual stems is the training data. Here we follow \cite{singsong} and use one of the state-of-the-art music source separation models \cite{demucs} to produce \textit{bass}, \textit{drum} and \textit{other} stems. Note that we intentionally exclude vocal stems from the present study, on one hand to avoid the complexity of the generation of coherent lyrics, and on the other hand due to legal constraints. 

\noindent \textbf{The proposed methods allow for several use-cases including:} (i) Generate music given a textual prompt and directly obtain 3 separated stems (\textit{bass}, \textit{drums} and \textit{other}); (ii) Generate the complementary stems (e.g. add the \textit{drums} and \textit{other} instruments on top of the \textit{bass}) given one or multiple stems (e.g. a \textit{bass}). Here again the generation may be controlled by means of an additional  text prompt; (iii) Remove and regenerate one or more of the three stems in an existing song; (iv) Modify the sound texture of the \textit{other} stem by regenerating its RVQ residuals while keeping its first stream fixed (see~\ref{editing}).

\noindent \textbf{Our contributions are as follows:} 
(i) We introduce \modelname  a variant of the  autoregressive text-to-music model MusicGen \cite{musicgen} that allows generating the three different stems that are used in the present study. The proposed method can generate all stems at once, or individual stems conditioned on a given musical sample; 
(ii) To prevent the possibility of cross-talk across the stems, we propose to use specialized compression models that are used to tokenize the individual stems; 
(iii) We evaluate on the text-to-music task and despite the additional complexity of the parallel generation of multiple stems, the proposed model is on par with its predecessor\cite{musicgen}. Additionally, on the unconditional generation task, it outperforms all the previous multi-stem generative models on objective and subjective metrics;
(iv) We introduce a particular conditioning approach that allows our model to perform stem editing (replace an existing stem) and stem by stem generation. Our evaluations show that our approach compares favorably to the previous open-source models that have been proposed for the task of stem editing.

\begin{figure*}[t]
    \centering
    \includegraphics[width=\linewidth]{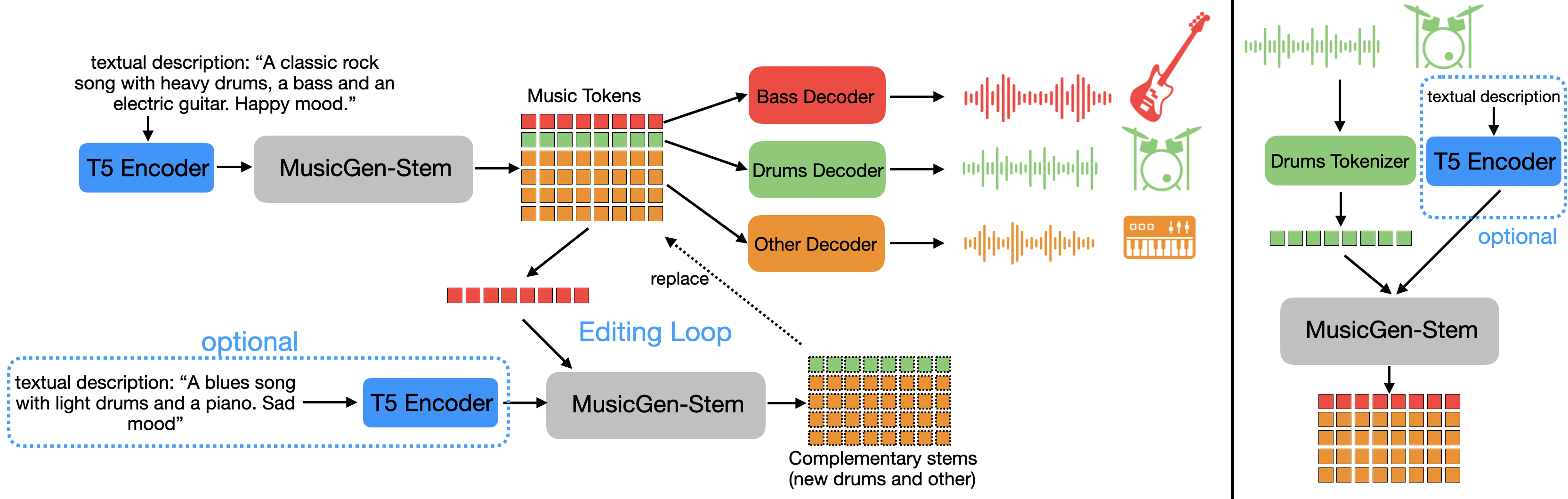}
    \caption{Three use-cases of our model: \textbf{(up)} \modelname can perform text-to-music generation and generates parallel streams of tokens representing the 3 stems (\textit{bass, drums and other}). \textbf{(down)} \modelname can also perform stem editing: given a subgroup of stems, the model can generate the complementary ones with an optional text prompt. \textbf{(right)} Given the waveform of one or multiple stems (that can be extracted from an existing song with Demucs), we tokenize them and \modelname can generate the missing stems with an optional text prompt. We can then decode them.}
    \label{fig:fig1}%
\end{figure*}

\section{Related Work}

\subsection{Music generation models}

One of the pioneering works in music generation was Wavenet~\cite{wavenet}, which introduced an autoregressive model for predicting the next sample in a quantized signal. This approach was initially inefficient during the sampling stage due to the high dimensionality of audio. 
The development of neural compression models~\cite{soundstream, encodec, dac} addressed this issue by representing one second of audio with a few hundreds audio tokens. Models such as MusicLM\cite{musiclm} and MusicGen~\cite{musicgen} leveraged this technique to build autoregressive models capable of generating coherent long-form music conditioned on text, audio, or even melodies. The authors in \cite{lemercier2024independence}, propose an improved optimization process in the training of the compression model to obtain better tokenization for autoregressive modeling. Other approaches~\cite{magnet, vampnet} have employed these compression models in a non-autoregressive manner, reducing latency but often at the expense of less convincing results.

Concurrently, diffusion models have established new standards in image generation~\cite{imagen}, with the advent of latent diffusion models~\cite{stablediff} specifically tackling the challenge of high-dimensional data. In the audio domain, methods like MusicLDM~\cite{musicldm}, AudioLDM~\cite{audioldm2} and Stable Audio~\cite{stableaudio} exemplify these advancements, with the latter being capable of generating up to 95 seconds of music. These methods operate on the intermediate representations of autoencoders, similar to compression models but without a quantization stage, thereby achieving higher fidelity.

\subsection{Music editing models}
Techniques of \textbf{zero-shot editing} such as DDPM inversion~\cite{zeroshot, ditto} and DDIM inversion~\cite{musicmagus} illustrate the potential of diffusion models to provide more flexibility and control in music generation. 
However, when attempting fine editing of a single instrument from a song, those approaches struggle to keep the rest of the track unchanged. This indicates that single stem models are not suited to the needs of real world artists.

\textbf{Test-time optimization} methods as well do not require training a model from scratch. For instance, textual inversion has been applied for diffusion~\cite{music_textual_inversion} and autoregressive~\cite{musicgen_style} models, where given a pretrained frozen text-to-music model and a batch of audio that share similarities (e.g. same artist, style or instruments), a ``pseudoword'' in the text embedding space representing these similarities is obtained by doing a gradient descent on the ``pseudoword'' by optimizing the loss of the model. Still, these inversion methods often result in artefacts in the generated audio.

However, discrete models based on quantized autoencoders lack of flexibility for editing. The autoregressive ones can regenerate the end of a song by using its beginning as a prompt but they cannot perform inpainting or stem editing which is crucial when one wants to modify specific sections of music without altering the entire piece.

To enable an autoregressive model to perform editing, instruction tuning can be done. For instance, in Instruct-MusicGen ~\cite{instruct_musicgen}, the authors fine-tune a pretrained MusicGen model with a source separation dataset and instructions in order to perform adding, removing, extracting and replacing instruments. This method is limited to 5 seconds generation and we observe that the task of stem editing often fails to keep the remaining stems unchanged because of the non separation of the instruments in the streams of the model.

\begin{figure*}[t]
    \centering
    \scalebox{1.0}{
    \includegraphics[width=1.0\textwidth]{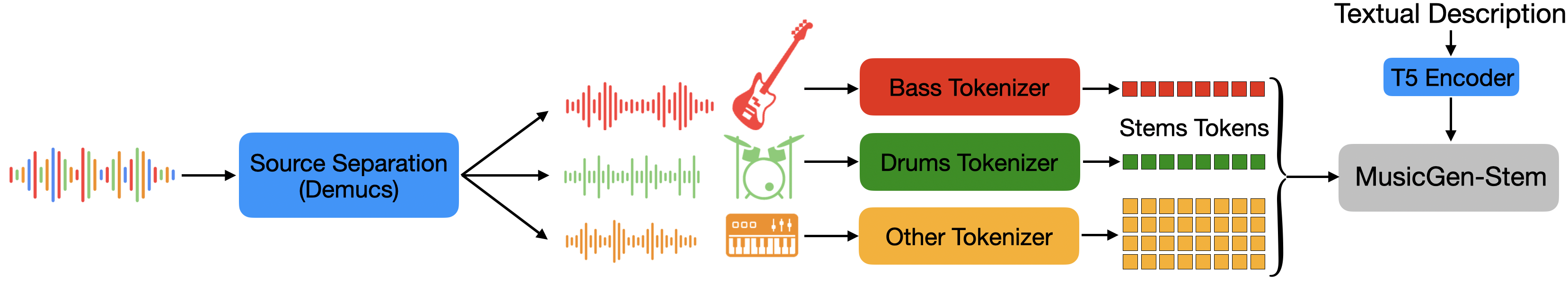}}
    \caption{Training pipeline. Given a song paired with its textual description, we process the song by using the source separation model Demucs and tokenize each stem with specific compression models. There is one stream of token for the \textit{bass} as well as the \textit{drums} and 4 streams of tokens for the \textit{other} instruments. Then, these tokens as well as the encoded textual description are fed into MusicGen-Stem's autoregressive transformer which is trained with a cross-entropy loss.} 
    \label{fig:training}
\end{figure*}

\subsection{Multi-stem music generation and editing}

In SingSong~\cite{singsong}, the authors perform vocal to accompaniment generation by conditioning an autoregressive MusicLM~\cite{musiclm} model with the coarse tokens of the compressed vocals stem. In StemGen~\cite{stemgen}, the authors sequentially generate music stem by stem with a masked model transformer. The drawbacks of their model is that it needs a first stem as an input and then generates music stem by stem which is compute intensive. In Jen-1 Composer~\cite{jen1composer} as well as in Multi-source diffusion models~\cite{multidiffusion_sep} (MSDM), the authors use a source separated dataset to train a diffusion model that outputs 4 stems in parallel. Jen-1 composer is a latent diffusion model whereas the second one is a diffusion model in the waveform space. In \cite{diffariff}, the authors train a latent diffusion model on a source separated dataset with a large set of instruments in an iterative manner. The only open-source model is MSDM~\cite{multidiffusion_sep}.

\section{Method}
In this section, we provide a detailed description of MusicGen-Stem. We start by describing the compression models. Next, we describe the auto-regressive sequential model. Lastly, we present the editing method. 
Fig.~\ref{fig:training} describes the training pipeline. 

\subsection{Compression models}
For each stem (\textit{drums, bass, other}), we train a compression model similar to EnCodec~\cite{encodec} that compresses 32kHz mono music into tokens at a rate of 50Hz. For the \textit{drums} and the \textit{bass} stem, we obtain good reconstruction quality with only a single quantization level. For the \textit{other} stem, we trained a model with 4 RVQ streams. Each of these specialized models is trained on our internal source separation dataset which consists of 3,000 professionally recorded songs. 
\subsection{Modeling and data preparation}

We train an autoregressive transformer for the task of text-to-music generation using cross entropy loss on 30 seconds audio segments at 50Hz. The 3 different stems are tokenized thanks to the 3 different compression models and their streams are concatenated and modelled in parallel. We use the medium (1.5B) architecture of MusicGen's transformer \cite{musicgen} and apply its delay pattern to the tokens. In our setup \textit{bass, drums} and the first RVQ \textit{other} stream are ``coarse'' tokens and in sync. Thus we only apply a delay on the 3 residual streams of the \textit{other} stem (see Fig.~\ref{fig:delay_pattern}).
Given the fact that we do not have a big dataset of labeled stem music, we train our model on the same data as in~\cite{musicgen} but we removed the songs that had vocals (15\% of our data) and use the last version of Demucs~\cite{demucs} to separate all the songs into 3 stems (\textit{bass, drums} and \textit{other}). 

\subsection{Editing}
\label{editing}
\begin{figure}[t]
    \centering
    \includegraphics[width=0.9\linewidth]{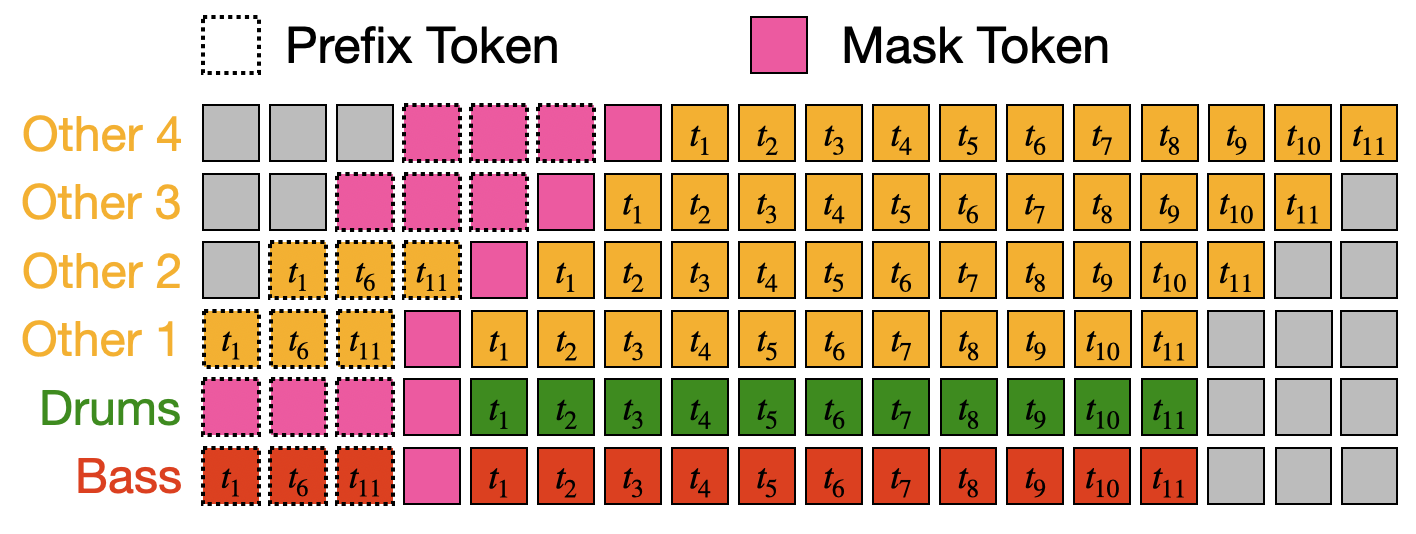}
    \caption{Training the editing task. Here the \textit{drums} and the 2 last streams of the \textit{other} stem are masked. The cross-entropy loss is computed on the tokens on the right of the masked tokens.}
    \label{fig:delay_pattern}%
    \vspace{-1.2em}
\end{figure}

At each training step we either train our model to perform text-to-music generation or editing with a 0.5 probability. To train the editing task, we take a sequence of 25 seconds (1,250 tokens), downsample it by a factor 5 (i.e. 10Hz) and use these 250 tokens as a prefix for the model. Then, we randomly sample 1 or 2 stems and mask the associated tokens in the prefix. If the \textit{other} stem is selected to be masked, we randomly choose to mask the streams in $\{4\}, \{4, 3\}, \{4, 3, 2\},$ or $\{4, 3, 2, 1\}$, forcing the model to learn to generate the details (the streams 2, 3, 4) of the \textit{other} stem given its first streams. We can see an example of a prefix on Fig.~\ref{fig:delay_pattern} where the \textit{drums} is masked as well as the streams 3 and 4 of the \textit{other} stem.

At inference time, we can 1) edit a song generated by the model (it is then already tokenized) 2) take an existing song separate its stems with Demucs and tokenize them 3) tokenize single stem music to be able to generate new stems. Then, we downsample to 10Hz this tokens sequence, we mask the desired stem and ask the model to continuous the generation in an autoregressive manner. During the autoregressive generation, we have the choice to force the unmasked streams to be exactly the same and only generate the masked streams or to let the model generate all of the streams. In the second case, we obtain a variation of the original unmasked stems (the model uses the downsampled prefix to reconstruct the stem). As well, the textual prompt let us control the generated new stems.

\vspace{-2mm}
\section{Experiments}

\subsection{Training details}

\modelname is trained for 400K steps with a batch size of 128. The data used for training include the internal MMI dataset that contains 10k high quality songs, ShutterStock and Pond5. We filtered the dataset to remove all songs containing vocals, resulting in a total of 17K hours of instrumental only data. We use AdamW optimizer with a learning rate of 1e-4.

\subsection{Metrics}
We evaluate the proposed method against state-of-the-art music generation models, considering both music generation and stem-editing setups. All the objective evaluations are performed on an internal test set of 534 songs for which we used Demucs to separate the stems.

\noindent
\textbf{Music generation:}  We use established objective and subjective metrics from the literature. Specifically the objective metrics used are: Frechet Audio Distance~\cite{fad}, the KL-divergence based metric introduced in~\cite{musicgen} and the CLAP score~\cite{clap} for text-to-music. For FAD we use the official TensorFlow implementation and pre-trained VGGish model. For subjective evaluations we follow the protocol proposed in MusicGen~\cite{musicgen} this consists in two studies, one for the overall quality of the samples (OVL) and one for the relation to text (REL).

\noindent
\textbf{Music editing:} To evaluate the editing performances of our model we perform two objective evaluations. The first one evaluates whether the rhythm matches between the original song and the new stem (BEAT). To do so, we use the beat tracking algorithm from \texttt{madmom}~\cite{madmom} both on the original song and on the generated stem. Then, we report the F-measure calculated with \texttt{mir\_eval}~\cite{mir_eval} using as reference the beats from the original song.
To evaluate the harmonic matching (HAR) between the \textit{bass} stem and the \textit{other} stem we use Chordino$^1$ to extract the chords played in the \textit{other} stem and use Pesto~\cite{pesto} to estimate the notes from the \textit{bass} line, we only keep the pitches predicted with a confidence greater than 0.75. We then compute the ratio of time steps where the \textit{bass} is playing a chord tone note~\cite{solfege}. For both metrics we zero out stems when the loudness was lower than $-35$dB.

In addition, we also conducted three subjective assessments of the editing process. In each assessment, we replaced one of the stems of the original song with a generated one. 
To ensure raters clearly ear the difference, we boost the generated stem to match the loudness of the rest of the track. We then set the overall loudness of this mix to -14dB.
Participants are told which instrument differs and are asked to rate the overall quality of the resulting songs. Every subjective study includes 40 samples that are each rated by at least 4 participants.

\def\thefootnote{1}\footnotetext{https://github.com/ohollo/chord-extractor}
\def\thefootnote{2}\footnotetext{https://github.com/gladia-research-group/multi-source-diffusion-models}

\subsection{Text-conditioned and unconditional music generation} 

In this section, we benchmark two families of generative models for music: text conditioned models and stem-level models. Note that only \modelname fits into both categories since MSDM does not handle text conditioning and MusicGen operates at the mixture level.
To evaluate MSDM~\cite{multidiffusion_sep} we use the official implementation$^2$. Since the original model (PT) is trained on the limited Slakh2100 dataset~\cite{slakh}, we include a version of this model trained on our dataset (RT).

\begin{table}[t!]
    \centering

    \resizebox{1.0\linewidth}{!}{\begin{tabular}{l|l l l | c c}
        Model & FAD $\downarrow$ & CLAP $\uparrow$  & KLD $\downarrow$  & REL $\uparrow$ & OVL $\uparrow$\\
        
        \midrule
        Ground Truth & $\times$ & 0.40 & $\times$ & 93.4 $\pm$0.7& 93.6 $\pm$0.5  \\
        \midrule      
       MusicGen$^*$ & 0.75  & 0.37 & \textbf{0.59} &  84.4 $\pm$1.0 & 86.7 $\pm$0.8 \\
       MusicGen-Stem$^*$ & \textbf{0.70} & \textbf{0.38} & 0.60 &\textbf{ 85.4} $\pm$0.7 & \textbf{87.0} $\pm$0.8 \\
    \midrule       
       MusicGen  & 2.13 & $\times$ & 1.02 & $\times$  & \textbf{85.0} $\pm$0.7 \\
       MSDM RT & 14.05 & $\times$ & 1.19 & $\times$  & 84.7 $\pm$0.8 \\
       MSDM PT & 7.61 & $\times$ & 1.48 & $\times$  & 80.9 $\pm$1.0 \\
       MusicGen-Stem & 2.15 & $\times$ & 1.04 & $\times$  & 83.8 $\pm$0.9 \\
    \end{tabular}}
    \caption{Comparisons of the different music generation models first in a text conditioned setup and then in an unconditional setup. Use of text conditioning is indicated with~$^*$.}
    
    \label{tab:ttm}
\end{table}

Objective and subjective metrics presented in TABLE~\ref{tab:ttm} suggest that MusicGen-Stem is on par with its predecessor on text-conditioned music generation. 
In the unconditional setup, our results suggest that MusicGen and MusicGen-Stem perform on par.
OVL scores shows that MSDM RT produces good quality outputs. However this model mostly generates similar songs, specifically ambient tracks with silent \textit{drums} and \textit{bass}. This limited diversity is reflected in a FAD score over 14.
\subsection{Text-conditioned and unconditional music editing}

We evaluate MusicGen-Stem on single stem music editing. The model is used to generate a coherent third stem in the context of two given stems. We compare it to both versions of MSDM and Instruct-MusicGen. Since the latter regenerates everything at the mixture level, it does not keep the input stems unchanged which prevents us to compute objective metrics.

\begin{table}[t!]
    \centering
    \resizebox{1.0\linewidth}{!}{
    \begin{tabular}{l|c c| c c c | c c c}
         & \multicolumn{2}{c|}{HAR $\uparrow$} & \multicolumn{3}{c|}{BEAT $\uparrow$} & \multicolumn{3}{c}{OVL $\uparrow$}\\   
      Edited stem & bass & other & bass & drums & other & bass & drums & other \\ 
        \midrule
        Ground Truth & 72\% & 72\% & 0.52 & 0.87 & 0.55 & 93.9 $\pm$0.7& 93.4 $\pm$0.7 & 93.5 $\pm$0.6 \\
        \midrule       
       MSDM RT & 48\% & 47\%  & 0.28 & 0.18 & \textbf{0.45} & 72.9 $\pm$1.6 & 54.0 $\pm$2.0 & 54.7 $\pm$1.6 \\
       MSDM PT & 31\% & 41\%  & 0.03 & 0.20 & 0.04 & 65.0 $\pm$2.0 & 54.6 $\pm$2.4 & 42.4 $\pm$2.7  \\
       Instruct-MusicGen & N/A & N/A  & N/A & N/A & N/A & 83.9 $\pm$0.9 & 51.4 $\pm$1.1 & 64.4 $\pm$1.1  \\
       MusicGen-Stem$^*$ & \textbf{66\%} & \textbf{68\%}  & 0.42 & \textbf{0.69} & 0.41 & 86.5 $\pm$0.8 &\textbf{86.8} $\pm$0.9 & \textbf{75.8} $\pm$1.7\\
       MusicGen-Stem & \textbf{66\%} & 67\%  & \textbf{0.46} & 0.67 & \textbf{0.45} & \textbf{86.7}$\pm$0.9 & 86.4 $\pm$0.8 & 72.6 $\pm$1.2\\
    \end{tabular}}
    \caption{Performances of the models on stem editing task. Use of text conditioning is indicated with~$^*$.}
    \label{tab:editing}
    \vspace{-2em}
\end{table}

Results from Table~\ref{tab:editing} indicate that MusicGen-Stem outperforms all evaluated baselines in stem editing, regardless of whether text conditioning is applied. Our model consistently generates stems that are more coherent with the overall track, both in terms of rhythm and pitch. Subjective evaluations further validate the superior editing performance of our model. While Instruct-MusicGen shows promising results in bass performance, it is constrained to 5-second audio clips and occasionally alters the song significantly.

\section{Conclusion}

We introduce a model that is capable of generating music conditioned on either text or instrument stems. MusicGen-Stem reaches comparable performance to the evaluated baselines when considering text-to-music generation, while allowing stem editing.
This makes it possible for musicians to iterate on their creations by being able to keep some parts at the instrument level. While \modelname is a step towards better control in music generation, it is still limited to 3 stems due to the lack of high quality dataset containing more than the classic \textit{bass}, \textit{drums} and \textit{other}. For future work we intend to increase the capacity of the \textit{bass} tokenizer that tends to create artefacts for higher pitch notes. We also want to have better control on the \textit{other} stem generation with refined conditioning like instrument embedding.

\newpage
\bibliographystyle{IEEEbib}
\bibliography{refs}
\end{document}